\documentclass[aps,prd,longbibliography,showpacs,twocolumn,superscriptaddress]{revtex4-2}
\usepackage{amsmath,amssymb,amsfonts,bm}
\usepackage{booktabs}
\usepackage{braket}
\usepackage{graphicx}
\usepackage{epstopdf}
\usepackage{dcolumn}
\usepackage{mathrsfs}
\usepackage{bbold}
\usepackage{dsfont}
\usepackage{float}
\usepackage{soul}
\usepackage[colorlinks=true,linkcolor=blue,citecolor=blue, urlcolor=blue,bookmarks=false]{hyperref}
\usepackage{amsbsy}
\usepackage{stmaryrd}
\usepackage{babel}
\usepackage{tgtermes}
\usepackage{multirow}
\usepackage[title,titletoc]{appendix}
\usepackage{mathptmx}
\DeclareMathAlphabet{\mathcal}{OMS}{cmsy}{m}{n}

\begin{document}

\title{Light-Tunable Giant Anomalous Hall Effect in the Flat-Band Magnetic Weyl Semimetal $\mathrm{AlFe_2O_4}$}

\author{Tingyan Chen}
\affiliation{Institute for Structure and Function $\&$ Department of Physics $\&$ Chongqing Key Laboratory for Strongly Coupled Physics, Chongqing University, Chongqing 400044, People's Republic of China}
\affiliation{Center of Quantum materials and devices, Chongqing University, Chongqing 400044, People's Republic of China}

\author{Shengpu Huang}
\affiliation{Institute for Structure and Function $\&$ Department of Physics $\&$ Chongqing Key Laboratory for Strongly Coupled Physics, Chongqing University, Chongqing 400044, People's Republic of China}
\affiliation{Center of Quantum materials and devices, Chongqing University, Chongqing 400044, People's Republic of China}

\author{Jing Fan}
\affiliation{Center for Computational Science and Engineering, Southern University of Science and Technology, Shenzhen 518055, China}

\author{Dong-Hui Xu}
\affiliation{Institute for Structure and Function $\&$ Department of Physics $\&$ Chongqing Key Laboratory for Strongly Coupled Physics, Chongqing University, Chongqing 400044, People's Republic of China}
\affiliation{Center of Quantum materials and devices, Chongqing University, Chongqing 400044, People's Republic of China}

\author{Rui Wang}
\email{rcwang@cqu.edu.cn}
\affiliation{Institute for Structure and Function $\&$ Department of Physics $\&$ Chongqing Key Laboratory for Strongly Coupled Physics, Chongqing University, Chongqing 400044, People's Republic of China}
\affiliation{Center of Quantum materials and devices, Chongqing University, Chongqing 400044, People's Republic of China}

\author{Da-Shuai Ma}
\email{mads@cqu.edu.cn}
\affiliation{Institute for Structure and Function $\&$ Department of Physics $\&$ Chongqing Key Laboratory for Strongly Coupled Physics, Chongqing University, Chongqing 400044, People's Republic of China}
\affiliation{Center of Quantum materials and devices, Chongqing University, Chongqing 400044, People's Republic of China}
	
\begin{abstract}
Achieving a giant anomalous Hall effect (AHE) and enabling its effective tuning are fundamental goals for topological spintronics. Magnetic Weyl semimetals hosting flat bands offer a promising route to maximize the AHE. However, while theoretical models are well-established, realistic material candidates remain scarce. 
Since the intrinsic anomalous Hall conductivity (AHC) is topologically dictated by the momentum separation ($\kappa$) between Weyl nodes, actively manipulating remains a key challenge.
Here, through comprehensive first-principles calculations, we establish the inverse spinel $\mathrm{AlFe_2O_4}$ as a realistic ferromagnetic half-metallic platform integrating three-dimensional flat bands and Weyl physics. Spin-orbit coupling induces a single pair of Weyl nodes, yielding a giant intrinsic AHC of $398\ \mathrm{S}\cdot\mathrm{cm}^{-1}$. By constructing a symmetry-constrained tight-binding model, we uncover a deterministic relationship between microscopic electronic couplings and the macroscopic AHE. Exploiting this via Floquet engineering with circularly polarized light, we demonstrate that the effective couplings are dynamically suppressed. This optical modulation controllably enlarges $\kappa$, shortens the topological Fermi arcs, and drives a dramatic, quantitative suppression of the AHC, providing a practical blueprint for ultrafast, light-controlled topological transport.
\end{abstract}
\maketitle

\section{\label{sec:level1}Introduction}
The anomalous Hall effect (AHE) remains one of the most captivating transport phenomena in condensed matter physics, fundamentally linking macroscopic charge currents to the microscopic quantum geometry of Bloch electrons~\cite{karplus1954hall,nagaosa2010anomalous}. Originating intrinsically from the momentum-space Berry curvature, which acts as a fictitious magnetic field induced by the interplay of time-reversal symmetry breaking and spin-orbit coupling (SOC), the AHE enables transverse current generation without the need for external magnetic fields~\cite{ye1999berry,yao2004first,jungwirth2002anomalous,xiao2010berry}. This dissipationless transport characteristic makes the AHE highly desirable for next-generation low-power spintronics and topological memory devices~\cite{moritz2008extraordinary,wang2020anomalous,lu2012ultrasensitive,dc2018room,zhang2022bipolar,puebla2020spintronic}. Recently, magnetic Weyl semimetals (WSMs) have emerged as a fertile playground for exploring the AHE, as their defining Weyl nodes act as singular magnetic monopoles of Berry curvature~\cite{xu2011chern,xu2015discovery,burkov2011weyl,burkov2014anomalous,zyuzin2016intrinsic,jiang2020magnetic}. In a magnetic WSM, the intrinsic AHC is topologically constrained by the distribution of nodes with opposite chirality. Theoretical studies have established that for a minimal WSM with a single pair of Weyl nodes, the intrinsic AHC peak is given as
\begin{equation}
    \sigma_{xy} = \frac{e^2}{2\pi\hbar} \mathcal{K}, \quad \text{with} \quad
    \mathcal{K} = 
    \begin{cases}
        \kappa, & \text{if } C_{k_c=0} = 1 \\
        1 - \kappa, & \text{if } C_{k_c=0} = 0
    \end{cases},
    \label{eq:AHC}
\end{equation}
where $C_{k_c=0}$ is the Chern number of the globally gapped $k_c=0$ plane separating Weyl nodes of opposite chiralities, and $\kappa$ (united by $2\pi/a$, with $a$ is the lattice constant) is the momentum separation between them~\cite{burkov2014anomalous}.

To effectively harness the AHE for next-generation devices, achieving a giant AHC is a fundamental prerequisite. While magnetic WSMs such as $\mathrm{Co_3Sn_2S_2}$ exhibit large AHEs~\cite{liu2018giant, wang2018large}, they suffer from an intrinsic limitation: their highly dispersive bands distribute multiple pairs of Weyl nodes across different energies, making it exceedingly difficult to align the Fermi level with the AHC peak. To circumvent this, three-dimensional (3D) flat band systems have been proposed as an ideal paradigm~\cite{mielke1991ferromagnetic,liu2014exotic}. The vanishing bandwidth in flat bands naturally pins the Fermi level, enabling the simultaneous harvesting of Berry curvature from multiple Weyl nodes to maximize the AHE \cite{jiang2021giant,he2024flat}. However, realistic material candidates that embody these 3D flat band models remain largely elusive. Furthermore, from a fundamental physics perspective, realizing a giant static AHE solves only half the problem; the ultimate goal lies in the dynamic control of transport properties. Actively manipulating the momentum separation ($\kappa$) between Weyl nodes allows for the real-time tuning of the Berry curvature distribution and the on-demand switching of macroscopic topological currents. Although static tuning strategies such as chemical doping~\cite{takahashi2009control} or strain engineering~\cite{gong2025intrinsic} have been explored, they fundamentally lack the reversibility and ultrafast speed required for practical device operation. In this context, Floquet engineering with periodic driving of a light field emerges as a fascinating avenue~\cite{bao2022light,oka2019floquet,zhan2024perspective,ezawa2013photoinduced,kitagawa2011transport,zhou2023pseudospin,zhan2023floquet,zhu2023floquet,oka2009photovoltaic}. As a time-periodic perturbation, the light field provides a robust optical mechanism to dynamically reconstruct electronic bands and precisely tailor $\kappa$. Despite recent progress in light-driven topological phase transitions~\cite{PhysRevB.94.121106,PhysRevLett.117.087402,PhysRevB.94.081103,hubener2017creating,PhysRevB.99.075121,PhysRevLett.123.206601,zou2025floquet,liu2018photoinduced,mciver2020light,huang2024controllable,fan2024circularly}, applying Floquet engineering to realistic 3D flat band materials to dynamically modulate the AHE remains an unexplored frontier. This raises a pivotal question: Beyond bridging the gap from theoretical models to physical realization, how can we uncover the microscopic origins of to dynamically manipulate macroscopic topological transport?

In this work, combining first-principles calculations and tight-binding models with Floquet engineering, we answer this pivotal question by demonstrating that the inverse spinel $\mathrm{AlFe_2O_4}$ serves as a ferromagnetic half-metallic material platform integrating 3D flat bands, Weyl physics, and a tunable AHC. In the absence of spin-orbit coupling (SOC), its electronic structure features a three-component fermion at the $\Gamma$ point, accompanied by quasi-flat bands located in close proximity to the Fermi level. The inclusion of SOC lifts this degeneracy, opening a gap at the $\Gamma$ point and inducing a single pair of Weyl nodes located at high-symmetry lines $-K \leftrightarrow \Gamma \leftrightarrow K$ with $\kappa \approx 0.292$. Consequently, the system hosts a giant anomalous Hall conductivity (AHC) peak of $\sim 398\ \mathrm{S}\cdot\mathrm{cm}^{-1}$ near the Fermi level. More interestingly, by constructing a symmetry-allowed minimum tight-binding model, we find that $\kappa$, which is linearly correlated with the AHC, is highly sensitive to nearest-neighbor couplings. Building on this mechanism, we employ Floquet engineering via circularly polarized light (CPL) to dynamically modulate the topological responses. This optical manipulation directly results in a significant shortening of the Fermi arcs on the (010) surface and a corresponding suppression of the AHC. These results establish $\mathrm{AlFe_2O_4}$ as a realistic material platform for investigating the interplay between 3D flat bands and Weyl physics, and provide a viable strategy for the optical tuning of topological transport.

\begin{figure}[t]
\includegraphics[width=\linewidth]{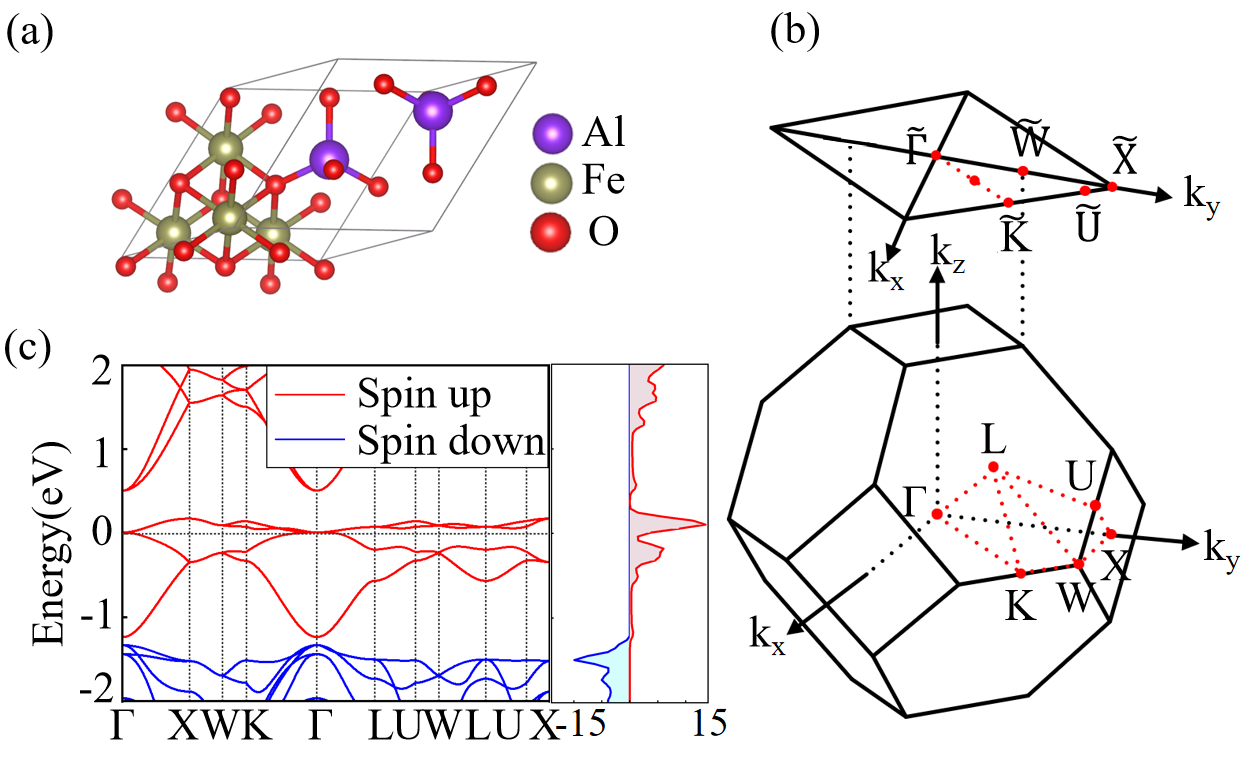}
\caption{\label{fig:fig1}
Crystal structure and spin-polarized electronic properties of the inverse spinel $\mathrm{AlFe_2O_4}$. 
(a) Schematic of the $\mathrm{AlFe_2O_4}$ crystal lattice, with Al, Fe, and O atoms represented by purple, dark yellow, and red spheres, respectively. 
(b) Bulk Brillouin zone (BZ) and its projection onto the (010) surface, with high-symmetry points labeled. 
(c) First-principles electronic band structure (left panel) and the corresponding density of states (DOS, right panel) calculated without spin-orbit coupling (SOC). The spin-up and spin-down channels are denoted by red and blue curves, respectively. The system exhibits a half-metallic nature with fully spin-polarized quasi-flat bands near the Fermi level, which give rise to a sharp peak in the DOS.}
\end{figure}

\section{\label{sec:level2}COMPUTATIONAL METHODS}
First-principles calculations were performed using the Vienna \textit{ab initio} simulation package (VASP) \cite{kresse1996efficient,kresse1996efficiency} within the framework of density functional theory (DFT) \cite{hohenberg1964inhomogeneous,kohn1965self}. The electron-ion interactions were described by the projector-augmented-wave (PAW) method \cite{blochl1994projector}. The generalized gradient approximation (GGA) with the Perdew-Burke-Ernzerhof (PBE) parametrization was employed for the exchange-correlation functional \cite{perdew1997generalized}. To properly account for the strong electron correlation effects of the Fe $3d$ electrons, the GGA+$U$ method \cite{liechtenstein1995density} was adopted with an effective Hubbard $U = 4$ eV, consistent with previous theoretical studies on $\mathrm{AlFe_2O_4}$. The experimental lattice constants were adopted \cite{otrokov2019prediction}, and the internal atomic positions were fully relaxed until the residual forces on each atom were less than $0.02\ \mathrm{eV/\text{\AA}}$. The energy cutoff for the plane-wave basis was set to 560 eV, and a $9 \times 9 \times 9$ Monkhorst-Pack k-point grid \cite{monkhorst1976special} was used for the Brillouin zone sampling. Spin-orbit coupling (SOC) was included self-consistently where specified.

To investigate the topological properties, we constructed a tight-binding Hamiltonian using the maximally localized Wannier functions (MLWFs) method \cite{marzari1997maximally,mostofi2008wannier90,qiao2018calculation}. Based on the orbital-projected band structure (fat-band analysis), the five $3d$ orbitals of the Fe atoms were selected as the initial projectors to construct the MLWFs \cite{marzari2012maximally}. Starting from this Wannier-based tight-binding (WFTB) Hamiltonian, we introduced the light-matter interaction via the Peierls substitution to derive a time-dependent Hamiltonian \cite{yan2016tunable,goldman2014periodically,bajpai2020robustness}. Here, the dipole approximation was applied, neglecting the spatial variation of the light field due to its much longer wavelength compared to the lattice constant \cite{bao2022light}. In the high-frequency (off-resonant) regime, where the Floquet bands are well separated by a large photon energy $\hbar\omega$, an effective static Hamiltonian can be derived using the high-frequency expansion \cite{yan2016tunable,goldman2014periodically}. We truncated this expansion to the first order to accurately describe the photon-dressed band structures. Finally, the topological surface states and Fermi arcs were calculated from this effective Floquet Hamiltonian using the WannierTools package \cite{wu2018wanniertools}.

\begin{figure}[t]
\includegraphics[width=\linewidth]{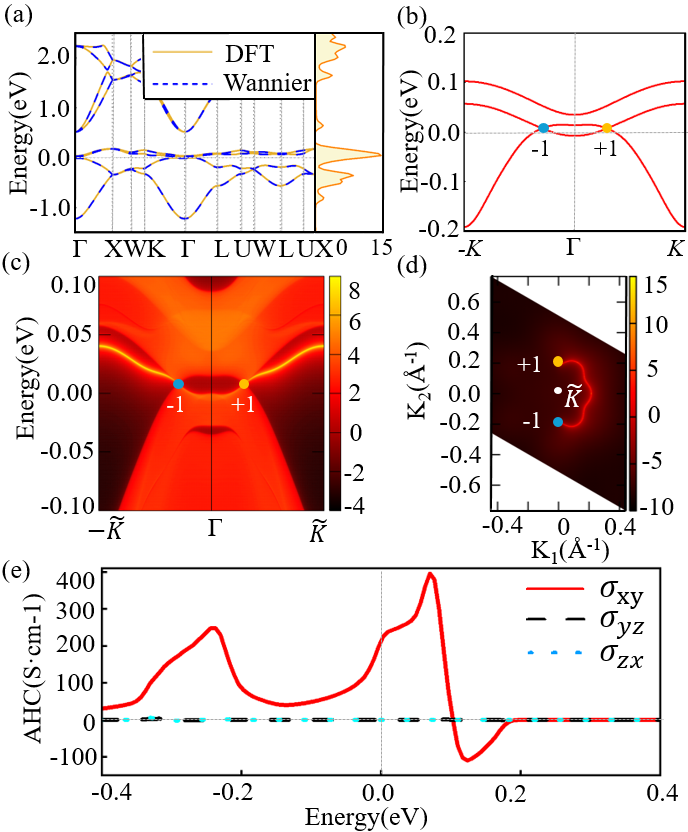}
\caption{\label{fig:fig2}
Topological electronic structure and anomalous Hall effect in $\mathrm{AlFe_2O_4}$ with SOC included. 
(a) Comparison of the band structure obtained from density functional theory (DFT, yellow solid curves) and the maximally localized Wannier function (MLWF) tight-binding model (blue dashed curves). The right panel shows the total DOS. 
(b) Magnified band dispersion along the $-K \leftrightarrow \Gamma \leftrightarrow K$ path, revealing the SOC-induced gap opening and the emergence of a single pair of Weyl nodes. The blue and yellow dots denote the Weyl nodes with chiral charges of $-1$ and $+1$, respectively. 
(c) Local density of states on the semi-infinite (010) surface. The bright red/yellow lines highlight the topological surface states connecting the projected Weyl nodes. 
(d) Constant-energy contour showing the Fermi arcs connecting the projected Weyl nodes of opposite chiralities on the (010) surface. 
(e) Calculated intrinsic anomalous Hall conductivity (AHC) as a function of energy. The solid red line represents the non-zero transverse component $\sigma_{xy}$, exhibiting a giant peak near the Fermi level, while the dashed black ($\sigma_{yz}$) and dotted blue ($\sigma_{zx}$) lines vanish due to symmetry constraints.}
\end{figure}

\section{\label{sec:level3}RESULTS AND DISCUSSION}
In this work, we focus on aluminum ferrite, $\mathrm{AlFe_2O_4}$, which crystallizes in an inverse spinel structure \cite{salcedo2023structural,hill1979systematics}. In a general normal spinel ($AB_2O_4$), the $A$ and $B$ cations strictly occupy tetrahedral and octahedral sites, respectively. However, an inverse spinel features a structural inversion where the $A$ cations and half of the $B$ cations co-occupy the octahedral network \cite{dong2017mechanical,verwey1947physical}. Following this configuration, as illustrated in Fig.~\ref{fig:fig1}(a), the $\mathrm{Fe^{3+}}$ ions in $\mathrm{AlFe_2O_4}$ occupy the tetrahedral A-sites (Wyckoff position $8a$), while the $\mathrm{Fe^{2+}}$ and $\mathrm{Al^{3+}}$ ions jointly occupy the octahedral B-sites (Wyckoff position $16d$) within a face-centered cubic (FCC) lattice (space group $Fd\bar{3}m$, No. 227, with $a = 5.67\ \text{\AA}$) \cite{verwey1947physical,jain2013commentary}. Crucially, the octahedral B-sites in the spinel framework form a three-dimensional pyrochlore network of corner-sharing tetrahedra. This unique geometric configuration is highly conducive to destructive quantum interference, naturally providing a structural basis for the emergence of electronic flat bands \cite{regnault2022catalogue}. Magnetically, $\mathrm{AlFe_2O_4}$ exhibits a robust ferromagnetic ground state. By aligning the spontaneous magnetization along the $[001]$ crystallographic axis, the system breaks time-reversal symmetry ($\boldsymbol{T}$) but preserves the $\boldsymbol{C}_{4z}$ rotational symmetry, which strictly constrains the topological transport properties of this system that would be discussed later.

To uncover the electronic properties of the $\mathrm{AlFe_2O_4}$ ferromagnetic phase, we first obtained the energy bands along the high-symmetry lines of the bulk Brillouin zone (BZ) illustrated in Fig. \ref{fig:fig1}(b). As shown in the DFT-calculated band structure in Fig. \ref{fig:fig1}(c), $\mathrm{AlFe_2O_4}$ without spin-orbit coupling (SOC) exhibits distinct half-metallic behavior. The electronic bands show complete spin polarization near the Fermi level, with the spin-up (red curves) and spin-down (blue curves) components clearly separated by a large exchange splitting. Consequently, the states around the Fermi energy ($E_F$) are exclusively dominated by the spin-up channel. Remarkably, the system hosts quasi-degenerate three-dimensional flat bands located in close proximity to the Fermi level. Specifically, two quasi-flat bands are degenerate along the $L \leftrightarrow \Gamma \leftrightarrow X \leftrightarrow W$ path. Originating from the geometric frustration of the pyrochlore network, these fully spin-polarized flat bands provide an exceptionally high density of states, as shown in the right panel of Fig. \ref{fig:fig1}(c). Furthermore, a three-component fermion emerges at the $\Gamma$ point, as the electronic bands at this wave vector host a three-dimensional irreducible representation of the little group. This high-fold degeneracy at the $\Gamma$ point, combined with the adjacent flat bands, sets the stage for strong topological responses of the magnetic Weyl semimetal $\mathrm{AlFe_2O_4}$ featuring quasi-flat bands.

Taking into account the non-negligible spin-orbit coupling (SOC) in this magnetic system, the low-energy electronic structure undergoes significant modifications. Specifically, as illustrated in Fig. \ref{fig:fig2}(a), the degenerate quasi-flat bands along the $L \leftrightarrow \Gamma \leftrightarrow X \leftrightarrow W$ path completely lose their degeneracy under SOC, even though they largely retain their flat dispersion profiles. Simultaneously, as shown in Fig. \ref{fig:fig2}(b), the high-fold degeneracy of the three-component fermion at the $\Gamma$ point is lifted, resulting in the opening of a distinct band gap. To carefully examine the topological consequences of this SOC-induced gap, we evaluated the energy difference between the highest valence band and the lowest conduction band. This analysis reveals gapless points corresponding to a pair of Weyl nodes located at the fractional coordinates of $(0, 0, \pm 0.146)$ along the $-K \leftrightarrow \Gamma \leftrightarrow K$ path. As highlighted in Fig. \ref{fig:fig2}(b), these nodes carry opposite chiral charges of $+1$ and $-1$, denoted by yellow and blue dots, respectively.

To visualize the topologically protected surface states, we calculated the surface state spectrum on the semi-infinite (010) surface, as presented in Fig. \ref{fig:fig2}(c). Furthermore, as shown in the corresponding constant-energy contour in Fig. \ref{fig:fig2}(d), the characteristic Fermi arcs are distinctly observed along the $\Gamma \leftrightarrow \widetilde{K}$ direction, connecting the projected Weyl nodes with opposite chiralities. This connectivity indicates that the two-dimensional momentum plane at $k_3 = 0$ is topologically trivial ($C_{k_3 = 0} = 0$), whereas the $k_3 = \pi$ plane is non-trivial ($C_{k_3 = \pi} = 1$). Beyond surface signatures, the topological non-triviality profoundly impacts the macroscopic transport. The intrinsic anomalous Hall conductivity (AHC) is evaluated by integrating the Berry curvature over the Brillouin zone (BZ) \cite{yao2004first}:
\begin{equation}\label{eq:ahc}
    \sigma_{ij} = -\frac{e^2}{\hbar} \int_{\text{BZ}} \frac{d^3\boldsymbol{k}}{(2\pi)^3} \sum_n f_n(\boldsymbol{k}) \Omega_{n,ij}(\boldsymbol{k}),
\end{equation}
where $f_n(\boldsymbol{k})$ is the Fermi-Dirac distribution and $\Omega_{n,ij}(\boldsymbol{k})$ is the Berry curvature tensor. Constrained by the $\boldsymbol{C}_{4z}$ rotational symmetry along the $[001]$ magnetization axis, the transverse AHC components vanish ($\sigma_{yz} = \sigma_{zx} = 0$). Consequently, the anomalous Hall response is entirely characterized by $\sigma_{xy}$. As explicitly plotted in Fig.~\ref{fig:fig2}(e), driven by the massive Berry curvature from the Weyl nodes and adjacent quasi-flat bands, the $\sigma_{xy}$ spectrum exhibits two pronounced peaks in the vicinity of the Fermi level. Specifically, a giant peak of $398\ \mathrm{S}\cdot\mathrm{cm}^{-1}$ emerges at $E = 0.072$ eV, coinciding with the density of states (DOS) peak shown in Fig.~\ref{fig:fig2}(a), while a secondary peak of $250\ \mathrm{S}\cdot\mathrm{cm}^{-1}$ is located at $E = -0.24$ eV. Moreover, even exactly at the Fermi level ($E_F$), the system maintains a substantial $\sigma_{xy}$ of $221\ \mathrm{S}\cdot\mathrm{cm}^{-1}$, highlighting the robust topological response of this flat-band magnetic Weyl system. This substantial value places $\mathrm{AlFe_2O_4}$ among the ranks of prominent topological materials exhibiting a giant AHE. It is comparable in magnitude to the benchmark $\mathrm{Co_3Sn_2S_2}$ ($505\ \mathrm{S}\cdot\mathrm{cm}^{-1}$) \cite{wang2018large}, and notably exceeds those reported for $\mathrm{Sn_2Nb_2O_7}$ ($100\ \mathrm{S}\cdot\mathrm{cm}^{-1}$) \cite{jiang2021giant}, $\mathrm{GdPtBi}$ ($60\ \mathrm{S}\cdot\mathrm{cm}^{-1}$) \cite{shekhar2018anomalous}, and $\mathrm{NdPtBi}$ ($14\ \mathrm{S}\cdot\mathrm{cm}^{-1}$) \cite{shekhar2018anomalous}.

\begin{figure}
\includegraphics[width=\linewidth]{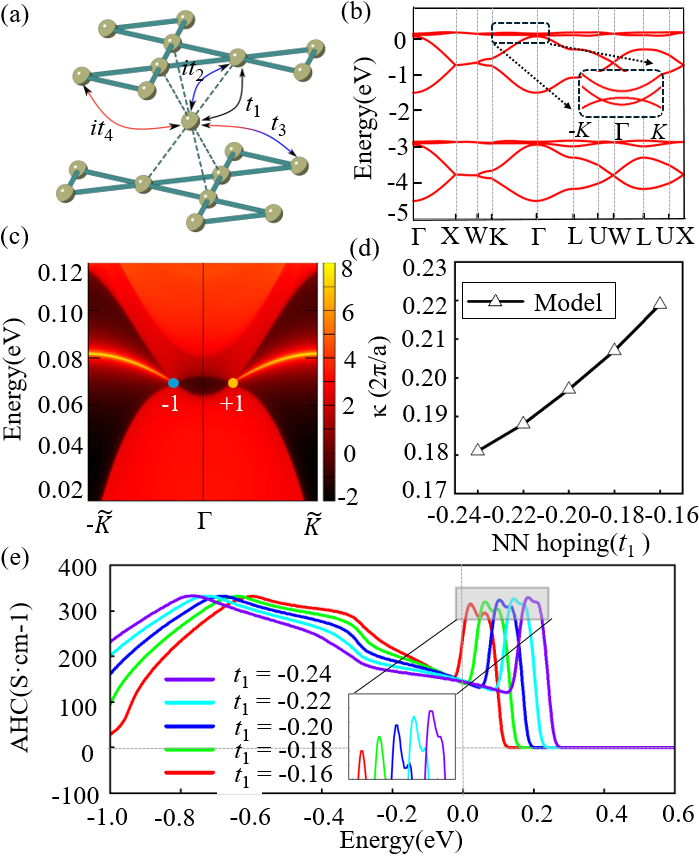}
\caption{\label{fig:fig3}Minimal tight-binding model and the microscopic origin of the tunable AHC. 
(a) Schematic of the symmetry-allowed effective lattice model on the pyrochlore sublattice (space group $Fd\bar{3}m$). The arrows indicate the considered hopping pathways, where $t_1$ represents the spin-independent nearest-neighbor (NN) hopping, and $t_2$, $t_3$, $t_4$ denote the SOC-induced spin-dependent hoppings. 
(b) Calculated bulk band structure of the effective model. The inset provides a magnified view of the band crossing along the $-K \leftrightarrow \Gamma \leftrightarrow K$ path, capturing the single pair of Weyl nodes. 
(c) Projected surface states derived from the minimal model, with the Weyl nodes marked by yellow ($+1$) and blue ($-1$) dots. 
(d) Evolution of the momentum separation $\kappa$ (in units of $2\pi/a$) between the Weyl nodes as a function of the NN hopping parameter $t_1$. 
(e) Dependence of the macroscopic AHC spectrum on $t_1$. Different colored curves correspond to varying magnitudes of $t_1$ (from $-0.24$ to $-0.16$), demonstrating that the AHC peak can be deterministically tuned by modulating the microscopic NN coupling.
}
\end{figure}

While achieving a giant static AHC is highly desirable, the ultimate goal for topological applications lies in its active manipulation. As established in Eq. (\ref{eq:AHC}), dynamically tuning the AHC requires uncovering the microscopic parameters that dictate the Weyl node separation $\kappa$. To reveal the dominant factors governing $\kappa$ in the ferromagnetic Weyl semimetal $\mathrm{AlFe_2O_4}$, we construct a symmetry-allowed effective minimal tight-binding (TB) model. Rather than employing a full multi-orbital description, we capture the essential band dispersion characteristics and topological features, particularly the quasi-flat bands and Weyl nodes, by defining an effective $s$-orbital model on the pyrochlore sublattice, which preserves the crucial crystalline symmetries of the space group $Fd\bar{3}m$ (No. 227). The schematic of this lattice model and the considered hopping pathways ($t_1\sim t_4$) are illustrated in Fig.~\ref{fig:fig3}(a). The primitive cell contains four equivalent sites (corresponding to the Wyckoff position $16c$), located at fractional coordinates $\boldsymbol{r}_1=(0,0,0)$, $\boldsymbol{r}_2=(1/2,0,0)$, $\boldsymbol{r}_3=(0,1/2,0)$, and $\boldsymbol{r}_4=(0,0,1/2)$. Incorporating the spin degree of freedom, the Hilbert space is spanned by an eight-component basis: $\{ |s_{\boldsymbol{r}_i}, \uparrow\rangle, |s_{\boldsymbol{r}_i}, \downarrow\rangle \}$ ($i=1,2,3,4$). In this sublattice-spin tensor product space, the effective Hamiltonian in $\boldsymbol{k}$-space is formulated as:
\begin{equation}\label{eq:Hamiltonian1}
    H(\boldsymbol{k}) = H_{NN} + H_{SOC} + \mu I_{4} \otimes s_z + u I_{8}.
\end{equation}
Here, the first term $H_{NN} = t_1\sum_{\langle i,j \rangle, \sigma} c_{i\sigma}^{\dagger}c_{j\sigma}$ represents the spin-independent nearest-neighbor (NN) hopping. The second term, $H_{SOC}$, introduces the effective spin-orbit coupling via symmetry-allowed spin-dependent hoppings with amplitudes of $t_2 \sim t_4$, which is essential for opening the topological gap. The third term, $\mu I_{4} \otimes s_z$, acts as a uniform ferromagnetic exchange field that lifts the spin degeneracy, where $I_4$ is the identity matrix in the sublattice space and $s_z$ is the Pauli matrix. Finally, $u I_8$ represents the on-site chemical potential. The detailed matrix elements of Eq. (\ref{eq:Hamiltonian1}) are provided in the Supplemental Material \cite{SM}.

Using a representative parameter set ($t_1=-0.2$, $t_2=-0.06$, $t_3=0.04$, $t_4=0.025$, $\mu=-1.5$, and $u=-1.8$, all in unit of eV), the calculated band structure successfully reproduces the essential features of the ferromagnetic Weyl semimetal $\mathrm{AlFe_2O_4}$ near the Fermi level, including quasi-flat bands and a single pair of Weyl nodes at $\boldsymbol{k}=(0,0,\pm 0.098)$, as displayed in Fig.~\ref{fig:fig3}(b). The corresponding surface states emanating from these Weyl points are clearly visualized in Fig.~\ref{fig:fig3}(c), confirming that our minimal model accurately captures the intrinsic topological properties of ferromagnetic Weyl semimetal $\mathrm{AlFe_2O_4}$. Crucially, this minimal model enables us to distill the dominant microscopic factor governing the Weyl node separation. Specifically, by modulating the NN hopping parameter $t_1$ from $-0.24$eV to $-0.16$eV, we observe that the momentum separation $\kappa$ (where $\kappa = 2k_0$) decreases significantly as the hopping magnitude $|t_1|$ increases, as depicted in Fig.~\ref{fig:fig3}(d). Governed by the topological relation $\sigma_{xy} = \frac{e^2}{2\pi\hbar} (1 - \kappa)$, this microscopic reduction in $\kappa$ directly dictates the macroscopic transport behavior, leading to a substantial increase in the AHC as the magnitude $|t_1|$ grows, as explicitly demonstrated in Fig.~\ref{fig:fig3}(e). Uncovering this deterministic, formula-backed relationship provides the crucial theoretical basis for experimentally feasible schemes to actively manipulate the AHC.

\begin{figure}[t]
\includegraphics[width=\linewidth]{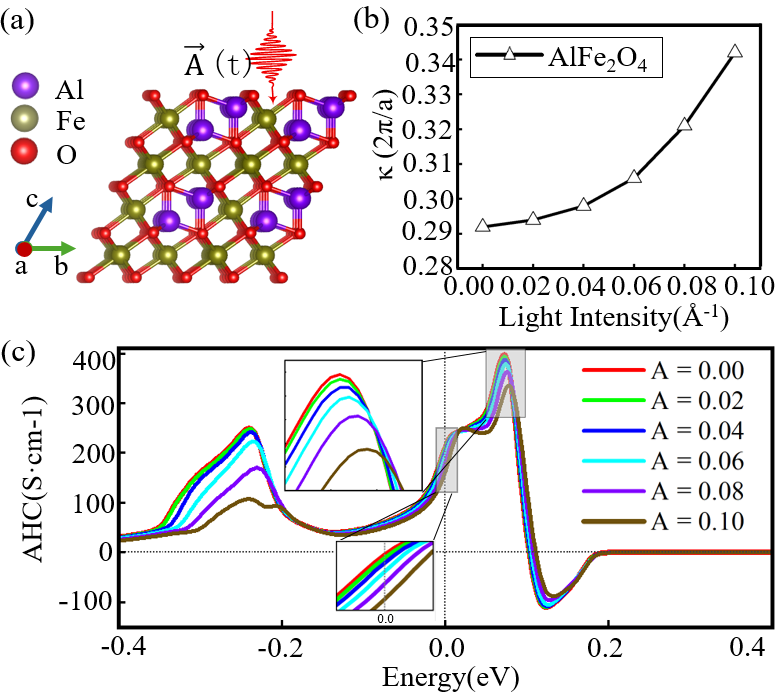}
\caption{\label{fig:fig4}
Floquet engineering of the macroscopic topological transport via circularly polarized light (CPL). 
(a) Schematic of the high-frequency CPL propagating along the $z$-axis (with vector potential $\boldsymbol{A}(t)$), which dynamically renormalizes the effective electronic couplings. 
(b) Momentum separation $\kappa$ (in units of $2\pi/a$) between Weyl nodes as a function of the dimensionless light intensity parameter $eA_0/\hbar$ (denoted as $A$ in the unit of $\text{\AA}^{-1}$). 
(c) Dynamic suppression of the AHC spectrum under varying light intensities. The different colored curves correspond to light intensities ranging from $A = 0.00$ to $0.10\ \text{\AA}^{-1}$. The two insets provide magnified views of the AHC evolution near the Fermi level ($E=0$) and at the primary peak ($E \approx 0.07\ \mathrm{eV}$), clearly illustrating the quantitative and controllable optical modulation of the macroscopic topological transport.
}
\end{figure}

Building on this deterministic relationship between the microscopic hopping and the macroscopic topological response, we now introduce Floquet engineering to dynamically manipulate the AHE. As schematically shown in Fig.~\ref{fig:fig4}(a), we apply high-frequency circularly polarized light (CPL) propagating along the $z$ axis, with the time-dependent vector potential defined as $\boldsymbol{A}(t) = A_0[\cos(\omega t), \sin(\omega t), 0]$, where $A_0$ and $\omega$ are the amplitude and frequency of the driving field, respectively. To preclude direct optical transitions and interactions between different Floquet sub-bands, a large photon energy of $\hbar\omega = 10\ \mathrm{eV}$ is adopted in our calculations.

The light-matter interaction is incorporated via the Peierls substitution. In the high-frequency regime, the photon-dressed Floquet states are effectively described by a static zeroth-order Hamiltonian, $H_{\mathrm{eff}}^{(0)}$. According to the Floquet-Magnus expansion, the effective hoppings are renormalized by the zeroth-order Bessel function of the first kind: $t_{ij}^{\mathrm{eff}} = t_{ij} J_0(\eta_{ij})$, where $\eta_{ij} = \frac{e A_0}{\hbar} \sqrt{\Delta x_{ij}^2 + \Delta y_{ij}^2}$ is the dimensionless light parameter, $eA_0/\hbar$ represents the light intensity, and $\sqrt{\Delta x_{ij}^2 + \Delta y_{ij}^2}$ denotes the projected distance between site $i$ and site $j$ onto the $x-y$ polarization plane. Notably, while the next-nearest-neighbor couplings are similarly renormalized by the light field, their contributions to the band reconstruction are secondary compared to those of the nearest-neighbor couplings. This stems from both their much smaller initial energy scales and their larger projected distances, which yield larger arguments $\eta_{ij}$ and consequently smaller values of the Bessel function $J_0(\eta_{ij})$. Therefore, the dynamic modulation of the topological properties is predominantly governed by the light-induced reduction of the nearest-neighbor couplings, perfectly mirroring the predictions from our static tight-binding analysis.

Guided by this mechanism, we systematically investigate the evolution of the Weyl nodes and macroscopic transport under varying light intensities ($eA_0/\hbar = 0.00$ to $0.10\ \text{\AA}^{-1}$). As shown in Fig.~\ref{fig:fig4}(b), the momentum separation $\kappa$ between the Weyl nodes monotonically increases with light intensity, significantly enlarging from $0.292$ at $eA_0/\hbar = 0.00$ to $0.343$ at $0.10\ \text{\AA}^{-1}$. Correspondingly, the topological surface states are strongly modulated, and the Fermi arcs on the (010) surface are effectively shortened (see Supplemental Material \cite{SM}). Beyond the momentum-space signatures, this dynamic reconstruction of the Berry curvature profoundly impacts the macroscopic transport. As illustrated in Fig.~\ref{fig:fig4}(c), governed by the relation $\sigma_{xy} \propto (1 - \kappa)$, the anomalous Hall conductivity exhibits a dramatic and controllable suppression in response to the incident light. Quantitatively, as the light intensity $eA_0/\hbar$ increases from $0.00$ to $0.10\ \text{\AA}^{-1}$, the intrinsic AHC at the Fermi level ($E=0$) is significantly reduced from $221$ to $163\ \mathrm{S}\cdot\mathrm{cm}^{-1}$. This light-induced suppression is robustly observed across the characteristic AHC peaks: the primary peak above the Fermi level decreases from $398\ \mathrm{S}\cdot\mathrm{cm}^{-1}$ (at $E=0.072\ \mathrm{eV}$) to $335\ \mathrm{S}\cdot\mathrm{cm}^{-1}$ (with a slight Floquet-induced shift to $E=0.078\ \mathrm{eV}$), while the secondary peak at $E=-0.24\ \mathrm{eV}$ experiences a massive reduction from $250$ to $108\ \mathrm{S}\cdot\mathrm{cm}^{-1}$. This direct and quantitative mapping from the optical driving field to the macroscopic topological response demonstrates a highly efficient, ultrafast, and fully reversible optical switching effect, offering an unprecedented degree of freedom for designing next-generation topological spintronic devices.

\section{\label{sec:level4}Summary and Outlook}
In summary, we have established the inverse spinel $\mathrm{AlFe_2O_4}$ as a realistic and ideal material platform that beautifully integrates three-dimensional flat bands, Weyl physics, and dynamically tunable anomalous Hall transport. Through comprehensive first-principles calculations and symmetry-constrained tight-binding modeling, we demonstrated that the interplay between the pyrochlore-network-induced flat bands and spin-orbit coupling generates a single pair of Weyl nodes, yielding a giant intrinsic AHC of $398\ \mathrm{S}\cdot\mathrm{cm}^{-1}$ near the Fermi level. Crucially, we uncovered a deterministic relationship between the microscopic electronic couplings and the macroscopic topological response. By applying high-frequency circularly polarized light, these effective couplings are dynamically suppressed via Floquet-Bessel renormalization. This optical modulation directly enlarges the momentum separation between Weyl nodes, shortens the topological Fermi arcs, and drives a dramatic, quantitative suppression of the AHC across multiple characteristic energy peaks.

From an experimental perspective, the physical realization of our theoretical proposal is highly feasible. First, high-quality $\mathrm{AlFe_2O_4}$ bulk single crystals have already been successfully synthesized in experiments~\cite{hill1979systematics,salcedo2023structural,dong2017mechanical}, providing a solid material foundation for our predictions. Second, the proposed Floquet engineering relies on light intensities (up to $eA_0/\hbar = 0.10\ \text{\AA}^{-1}$) that are readily achievable in current ultrafast laser experiments\cite{bao2022light,RevModPhys.96.015003}. The predicted dynamic band reconstruction, specifically the evolution of the Weyl nodes and the shortening of the Fermi arcs, can be directly visualized using time- and angle-resolved photoemission spectroscopy (TrARPES)\cite{bao2022light,RevModPhys.96.015003}. Concurrently, the ultrafast suppression of the macroscopic AHC can be precisely detected through  ultrafast terahertz transport measurements\cite{mciver2020light,PhysRevResearch.2.043408,kitagawa2011transport}. Ultimately, our findings not only bridge the gap between abstract flat-band Weyl models and realistic materials but also provide a practical blueprint for developing ultrafast, light-controlled topological spintronic devices.

\section*{ACKNOWLEDGMENTS}
\vspace{-0.8em}
This work was supported by 
National Key Research and Development Program of the Ministry of Science and Technology of China (Grant No.~2025YFA1411303), 
the National Natural Science Foundation of China (NSFC, Grants No.~12595330, No.~92365101, No.~12474151, No.~92565103, No.~12547101), 
the Natural Science Foundation of Chongqing (Grants No.~2023NSCQ-JQX0024), 
Beijing National Laboratory for Condensed Matter Physics (Grant No.~2024BNLCMPKF015 and No.~2024BNLCMPKF025), 
Open Research Fund of State Key Laboratory of Quantum Functional Materials (NO.~QFM2025KF002), and
the Postdoctoral Fellowship Program of CPSF (GZC20252254)
and the Fundamental Research Funds for the Central Universities (Grant No. 2025CDJIAISYB-032).
	
\bibliography{Manuscript}

\end{document}